\documentclass[aps,prb,twocolumn,groupedaddress]{revtex4}

\usepackage{amsmath}
\usepackage{amssymb}
\usepackage{graphicx}
\usepackage{braket}

\newcommand{\be}{\begin{equation}}
\newcommand{\ee}{\end{equation}}

\newcommand{\beq}{\begin{eqnarray}}
\newcommand{\eeq}{\end{eqnarray}}
\newcommand{\ba}{\begin{array}}
\newcommand{\ea}{\end{array}}

\def\H1{\widehat{H}_1}

\begin{document}

\title{Pseudospin excitations in coaxial nanotubes.}
\author{B. Scharf, J. Fabian, and A. Matos-Abiague}
\affiliation{Institute for Theoretical Physics, University of Regensburg, 93040 Regensburg, Germany}
\date{\today}

\begin{abstract}
In a 2DEG confined to two coaxial tubes the `tube degree of freedom' can be described in terms of pseudospin-1/2 dynamics. The presence of tunneling between the two tubes leads to a collective oscillation known as pseudospin resonance. We employ perturbation theory to examine the dependence of the frequency of this mode with respect to a coaxial magnetic field for the case of small intertube distances. Coulomb interaction leads to a shift of the resonance frequency and to a finite lifetime of the pseudospin excitations. The presence of the coaxial magnetic field gives rise to pronounced peaks in the shift of the resonance frequency. For large magnetic fields this shift vanishes due to the effects of Zeeman splitting. Finally, an expression for the linewidth of the resonance is derived. Numerical analysis of this expression suggests that the linewidth strongly depends on the coaxial magnetic field, which leads to several peaks of the linewidth as well as regions where damping is almost completely suppressed.
\end{abstract}


\maketitle

\section{Introduction}
\label{Intro}

Due to the Coulomb interaction between the charge carriers, collective excitations called plasmons may result as the electromagnetic response in a solid-state structure. These excitations have been extensively investigated in bulk and low-dimensional systems from both theoretical and experimental points of view.\cite{Pines:1964,Fetter:1980,Kushwaha2001:SSR,Giuliani:2005,Schuller:2006} However, the interest on plasma excitations has been renewed due to the experimental possibility of tailoring interactions such as spin-orbit coupling, which may affect the properties of the plasmons.\cite{Wang2005:PRB,Gumbs2005:PRB,Kushwaha2006:PRB,Pletyukhov2006:PRB,Badalyan2009:PRB} The recent ability of producing novel two-dimensional systems such as graphene has also motivated new investigations of the plasmon dispersion.\cite{Hwang2007:PRB,Polini2008:PRB,Gangadharaiah2008:PRL,Eberlein2008:PRB}

Interestingly, collective excitations can also emerge in systems involving two spatially separated two-dimensional electron gases which couple to each other through the Coulomb interaction.\cite{DasSarma1981:PRB,Jain1987:PRB,DasSarma1998:PRL,Kainth1999:PRB,Hu2001:PRB,Holland2002:PRB,Bootsmann2003:PRB} A typical example is the excitation of different plasmon modes in bilayer systems, where two quasi-2D electron systems (each with only the lowest subband being occupied) are separated by a potential barrier. Even when the potential barrier is large and the tunneling is largely suppressed, the interlayer Coulomb interaction may couple the two quasi-2D electron systems if the interlayer separation (barrier width) is small enough. In such a case there are intralayer plasmon excitations in which the electrons in one of the layers may collectively oscillate in phase (optic plasmon mode) or out of phase (acoustic plasmon mode) with the electron oscillations in the neighboring layer.\cite{DasSarma1981:PRB,Jain1987:PRB} By decreasing the barrier, the tunneling becomes relevant and splits the single subband in each quasi-2D electron system. As a result new excitations consisting of interlayer collective charge oscillations (intersubband or transverse plasmons) appear.\cite{DasSarma1998:PRL,Kainth1999:PRB,Hu2001:PRB,Holland2002:PRB,Bootsmann2003:PRB} Such interlayer collective oscillations have recently been re-interpreted as pseudospin excitations.\cite{Abedinpour2007:PRL}

Within the pseudospin approach, the electrons in one of the layers are assigned one pseudospin state and the electrons in the other layer the opposite pseudospin.\cite{MacDonald1990:PRL,Abedinpour2007:PRL} The interlayer excitations can be regarded as pseudospin excitations mediated by the tunneling strength, which acts on the pseudospins as an effective magnetic field. Thus, analogous to the conventional ferromagnetic resonance in magnetized materials whose electron spins are manipulated by an external magnetic field, the tunneling (effective magnetic field) in the bilayer system leads to a pseudospin resonance describing the interlayer collective mode (intersubband plasmon).\cite{DasSarma1998:PRL,Kainth1999:PRB,Hu2001:PRB,Holland2002:PRB} Furthermore, since the pseudospin degree of freedom is an analog to the real spin, new \emph{pseudospintronic} devices could be realized by means of controlled pseudospin manipulation, in close analogy with the control of real spin in spintronics applications.\cite{Zutic2004:RMP,Fabian2007:APS} In particular, a pseudospintronic device based on semiconductor bilayers has theoretically been suggested\cite{Abedinpour2007:PRL} as the analog to the conventional spin-transfer oscillator.\cite{Kiselev2003:N,Rippard2004:PRL,Tulapurkar2005:N}

Due to many-body effects the magnetization dynamics in magnetized systems are affected by the so-called Gilbert damping.\cite{Gilbert2004:IEEETM,Hankiewicz2007:PRB} Such effects also have their analog in bilayer systems, where the Coulomb electron-electron interaction produces a shifting of the pseudospin resonance frequency and leads to a finite lifetime of the excitations (that is, to damping).\cite{Abedinpour2007:PRL} Therefore, the investigation of the pseudospin excitations is of relevance for understanding the nature of correlations in bilayer-like systems.

Another interesting issue is the investigation of the pseudospin resonance in systems with more exotic geometries. Nowadays techniques allow for the realization of a wide range of possible geometries by using semiconductors or carbon based materials, for example. From this point of view, the phenomenon of pseudospin resonance offers the possibility of investigating many-body effects under different geometric configurations. In what follows we focus our discussion on the case of generic coaxial nanotubes which can be experimentally realized from a variety of materials including metals, metal-oxides, carbon, and semiconductors.\cite{Peng2009:Nano,Zhang2001:JMR,Hu2003:AM,Iijima1991:N,Dresselhaus:1987,Prinz2000:PE,Mendach2004:PE,Li2008:JPDAP} Coaxial nanotubes are particularly interesting systems for pseudospintronics, since they exhibit both the bilayer-like behavior of pseudospin excitations and the interplay between many-body and coherent effects which become apparent when a coaxial magnetic flux pierces the system. In such a case, as will be shown below, pseudospin resonance can be induced not only by an external electric field but also by fluctuations of the coaxial magnetic field.

Although there have been some investigations on plasma excitations in coaxial nanotubes\cite{Yannouleas1994:PRB,Lin1993:PRB} these studies were limited to the non-tunneling regime treated within the random-phase approximation (RPA). Here, we consider the possibility of tunneling (and therefore of pseudospin excitations) between the inner and outer tubes. Furthermore, we use a perturbative scheme recently proposed in Ref.~\onlinecite{Abedinpour2007:PRL} which appears to be superior to the RPA.

The paper is organized as follows: In Sec.~\ref{TM}, the pseudospin degree of freedom and the model Hamiltonian of the
system are introduced. To construct the Hamiltonian, we start with a single-particle Hamiltonian taking into account tunneling effects and analyze the energy spectrum of the resulting, before we include the electron-electron interaction. In Sec.~\ref{PR}, the pseudospin resonance and its corresponding response function are introduced. The perturbation scheme is set up in Sec.~\ref{PT}. The scheme is then applied in Secs.~\ref{PRF} and \ref{LW} to calculate the resonance frequency and the corresponding linewidth, respectively. The paper is concluded by a small summary.

\section{Theoretical Model}
\label{TM}

\begin{figure}
\includegraphics*[width=8cm]{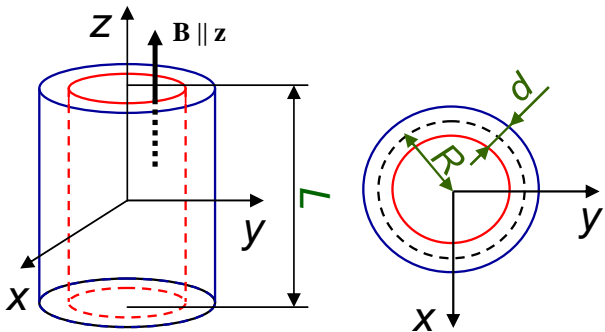}
\caption{(color online). Schematic of two coaxial tubes in the presence of an external magnetic field $\mathbf{B}$ along their axis. The length of the tubes and the intertube distance are denoted by $L$ and $d$, respectively. The radius of the inner (outer) tube is given by $R-d/2$ ($R+d/2$).} \label{fig:Model}
\end{figure}

The system (see Fig.~\ref{fig:Model}) consists of two coaxial tubes, which have the radii $R\pm d/2$ and the length $L\gg R\gg d$. A 2DEG is confined to the surface of each cylinder.

To approximate the band structure of actual solids, we assign an effective mass $m$ to the electrons. Similarly, the Coulomb interaction contains the background dielectric constant $\tilde{\epsilon}$, which is defined by
$\tilde{\epsilon}=\epsilon_{r}\epsilon_{0}$, where $\epsilon_{r}$ is the relative dielectric constant of the solid. The
corresponding Bohr radius in this material is then defined as $a_{B}=4\pi\tilde{\epsilon}\hbar^2/me^2$. The interior of the
cylindrical system is threaded by a static, coaxial magnetic field $\mathbf{B}$, which acts as a control parameter. To describe paramagnetic effects induced by the magnetic field, that is, Zeeman splitting, the $g$ factor of the material has to be introduced. In this model electrons can either be located on the outer or inner tube.

\subsection{Single-particle approximation}

In the absence of tunneling (tunneling effects will be included later on) the wave functions are localized at the tubes and can be approximated as
\begin{equation}
\Psi_{\rm{out/in}}\left(\textbf{r}\right)=\Phi_{\rm{out/in}}\left(\varphi,z\right)f_{\rm{out/in}}(\rho),
\end{equation}
where we have used cylindrical coordinates (see Fig.\ref{fig:Model}). The radial localization is characterized by
the function
\begin{equation}
\left|f_{\rm{out/in}}(\rho)\right|^2=\frac{\delta\left(\rho-R\mp\frac{d}{2}\right)}{\rho},
\end{equation}
where $\delta(x)$ represents the Dirac-delta function and the radial quantum number denotes whether the electron is located on the outer (\textit{out}) or inner (\textit{in}) tube. The longitudinal and azimuthal motions as well as the physical spin of the electrons are described by the spinors $\Phi_{\rm{out/in}}\left(\varphi,z\right)$. Apart from the physical spin, the radial motion can be reinterpreted as an effective two-level system, which we describe in terms of pseudospin-$1/2$ dynamics. We then replace the functions $f_{\rm{out/in}}$ by pseudospinors whose labels $\Uparrow$ and $\Downarrow$ correspond to the wave function localization in the outer and inner  tubes, respectively.

In the absence of tunneling the \textit{pseudospin} system is described by the single-particle Hamiltonian
\begin{equation}
\hat{H}_{0}=\sum\limits_{n,k,\sigma,\tilde{\Lambda}}\epsilon_{n,k,\sigma,\tilde{\Lambda}}\hat{a}_{n,k,\sigma,\tilde{\Lambda}}^{\dagger}\hat{a}_{n,k,\sigma,\tilde{\Lambda}},
\label{HamiltonianNoTunneling2nd}
\end{equation}
where $\hat{a}_{n,k,\sigma,\tilde{\Lambda}}$ ($\hat{a}_{n,k,\sigma,\tilde{\Lambda}}^{\dagger}$) is the operator of annihilation (creation) of a particle with angular momentum $n$, momentum along the $z$-axis $k$, spin $\sigma=\uparrow,\downarrow$ and pseudospin $\tilde{\Lambda}=\Uparrow,\Downarrow$. The single-particle eigenenergies are given by
\begin{equation}
\begin{aligned}
\epsilon_{n,k,\sigma,\tilde{\Lambda}}=&\frac{\hbar^2k^2}{2m}+\frac{g\hbar^2\Phi}{2m_eR^2\Phi_{0}}\sigma\\
&+\frac{\hbar^2}{2m\left(R+\tilde{\Lambda}\frac{d}{2}\right)^2}\left[n+\frac{\Phi}{\Phi_{0}}\left(1+\tilde{\Lambda}\frac{d}{2R}\right)^2\right]^2.\label{speettwt}
\end{aligned}
\end{equation}
Here, and in what follows, we use capital and lowercase Greek characters for denoting pseudospin ($\Uparrow,\Downarrow$) and
physical spin ($\uparrow,\downarrow$), respectively. We have also introduced the average magnetic flux $\Phi=\pi BR^2$, the fluxon $\Phi_{0}=h/e$, and the free electron mass $m_e$.

We now consider the possibility of uniform tunneling between both tubes. The tunneling amplitude $\Delta$ is assumed to be independent of the external magnetic field. Such an approximation is reasonable in systems in which the confinement is stronger than the cyclotron effects. The tunneling Hamiltonian,
\begin{equation}
\hat{H}_{t}=-\frac{\Delta}{2}\hat{\sigma}_{x},\label{Tunneling2nd}
\end{equation}
with $\hat{\sigma}_x$ as the corresponding pseudospin Pauli matrix, allows for the coupling of states localized in different tubes but with the same values of the quantum numbers $\sigma$ and $n$, which is consistent with the conservation of spin and angular momentum during the tunneling.

The form of $\hat{H}_{t}$ [see Eq.~(\ref{Tunneling2nd})] makes it clear that the tunneling amplitude can be interpreted as the pseudospin analog to a magnetic field in the $x$-direction. This term arises due to the overlap of the actual radial wave functions (which in reality are not as perfectly localized as our $\delta$-like model functions).

In analogy to the spin operator we can introduce the pseudospin vector operator $\hat{\mathbf{S}}$ whose $x$-component,
\begin{equation}\label{sx-def}
\hat{S}_{x}=\frac{1}{2}\sum\limits_{n,k,\sigma}\left(\hat{a}^{\dagger}_{n,k,\sigma,\Uparrow}\hat{a}_{n,k,\sigma,\Downarrow}+
\hat{a}^{\dagger}_{n,k,\sigma,\Downarrow}\hat{a}_{n,k,\sigma,\Uparrow}\right),
\end{equation}
characterizes the tunneling between the tubes. Indeed, the tunneling Hamiltonian in Eq.~(\ref{Tunneling2nd}) can be rewritten in second quantization as
\begin{equation}
\hat{H}_{t}=-\Delta\hat{S}_{x},\label{Tunneling2nd-v2}.
\end{equation}
The $y$-component of the pseudospin operator is given by
\begin{equation}\label{sy-def}
\hat{S}_{y}=\frac{i}{2}\sum\limits_{n,k,\sigma}\left(\hat{a}^{\dagger}_{n,k,\sigma,\Downarrow}\hat{a}_{n,k,\sigma,\Uparrow}-\hat{a}^{\dagger}_{n,k,\sigma,\Uparrow}\hat{a}_{n,k,\sigma,\Downarrow}\right)
\end{equation}
and measures the tunneling current flowing between the two tubes, while the $z$-component,
\begin{equation}
\begin{aligned}
\hat{S}_{z}&=\frac{1}{2}\sum\limits_{n,k,\sigma}\left(\hat{a}^{\dagger}_{n,k,\sigma,\Uparrow}
\hat{a}_{n,k,\sigma,\Uparrow}-\hat{a}^{\dagger}_{n,k,\sigma,\Downarrow}\hat{a}_{n,k,\sigma,\Downarrow}\right)\\
&=\hat{N}_{\Uparrow}-\hat{N}_{\Downarrow},
\end{aligned}\label{sz-def}
\end{equation}
measures the charge accumulation in the tubes. In Eq.~(\ref{sz-def}) $\hat{N}_{\Uparrow}$ and $\hat{N}_{\Downarrow}$
denote the number-of-particles operators in the outer ($\Uparrow$) and inner ($\Downarrow$) tubes, respectively.

\begin{figure}
\includegraphics*[width=8cm]{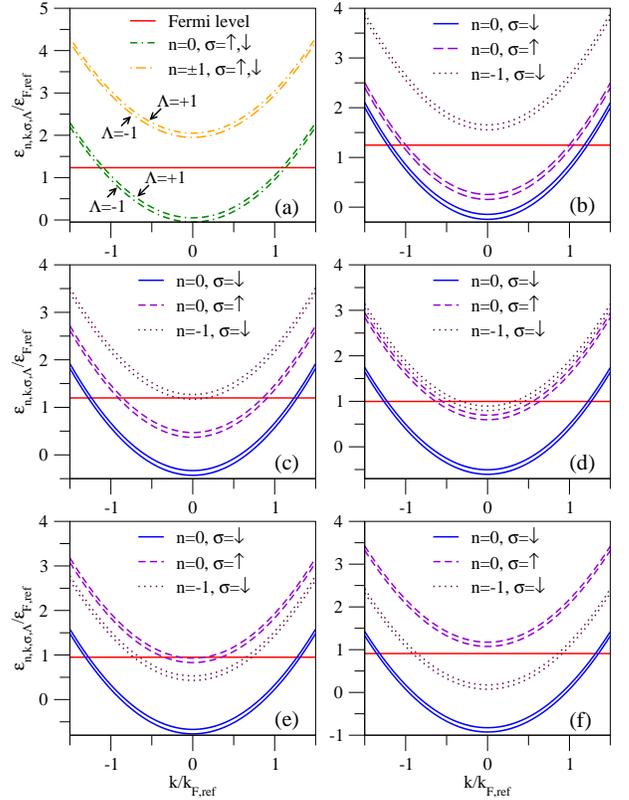}
\caption{(color online) Single-particle energy spectrum in the presence of tunneling for different values of the magnetic flux [(a): $\Phi=0$, (b): $\Phi=0.05\Phi_0$, (c): $\Phi=0.10\Phi_0$, (d): $\Phi=0.15\Phi_0$, (e): $\Phi=0.20\Phi_0$, (f): $\Phi=0.25\Phi_0$] with parameters $\Delta=0.1\epsilon_{\rm{F,ref}}$, $d=0.1R$, $g=2$, $r_{s}=1$, $R=0.5a_{B}r_{s}$, and $m=m_e$. The quantum numbers $n$, $\sigma=\uparrow,\downarrow$, and $\Lambda=\pm 1$ refer to the angular momentum, physical spin, and pseudospin, respectively. In all the cases $\epsilon_{n,k,\sigma,1}>\epsilon_{n,k,\sigma,-1}$, as explicitly indicated in (a).} \label{fig:spectrum}
\end{figure}

The single-particle Hamiltonian including the tunneling effects is given by $\hat{H}_{0}+\hat{H}_{t}$. Since $\hat{S}_{z}$ and $\hat{S}_{x}$ do not commute, the quantum number $\tilde{\Lambda}=\Uparrow,\Downarrow$, which classifies the
eigenvectors of $\hat{S}_{z}$, has to be replaced by another quantum number $\Lambda=\pm 1$ describing the pseudospin degree of freedom in the presence of tunneling, whereas the momentum along the $z$-axis as well as the angular momentum and spin projections on the $z$-axis remain good quantum numbers. The new pseudospinors with pseudospin quantum number $\Lambda$ are linear combinations of the pseudospinors with $\tilde{\Lambda}=\Uparrow,\Downarrow$ and describe \textit{bonding} ($\Lambda=-1$) and \textit{antibonding} ($\Lambda=+1$) states. In the basis of $\Lambda$-pseudospinors, the Hamiltonian $\hat{H}_{0}+\hat{H}_{t}$ becomes diagonal with the energy spectrum given by
\begin{equation}
\epsilon_{n,k,\sigma,\Lambda}=\frac{\epsilon_{n,k,\sigma,\Uparrow}+\epsilon_{n,k,\sigma,\Downarrow}}{2}+\frac{\Lambda}{2}\Delta_{n},\label{speettt}
\end{equation}
where
\begin{equation}
\Delta_{n}\equiv\sqrt{\left(\epsilon_{n,k,\sigma,\Downarrow}-\epsilon_{n,k,\sigma,\Uparrow}\right)^2+\Delta^2}.
\end{equation}
Here, $\epsilon_{n,k,\sigma,\Uparrow}$ and $\epsilon_{n,k,\sigma,\Downarrow}$ are the eigenenergies of $\hat{H}_{0}$ which are given by Eq.~(\ref{speettwt}).

From Eq.~(\ref{speettt}) one can see that the energy spectrum consists of one-dimensional subbands labelled by the quantum
numbers $n$, $\sigma$, and $\Lambda$. The evolution of the lowest subbands with increasing magnetic flux is shown in
Fig.~\ref{fig:spectrum}. The quantity of reference in this plot is $\epsilon_{\rm{F,ref}}$, the Fermi energy of a flat 2DEG with the same density $n_{e}$ as the cylindrical 2DEG considered here, and $k_{\rm{F,ref}}=\sqrt{2m\epsilon_{\rm{F,ref}}/\hbar^2}$. We assumed a model system with parameters $\Delta=0.1\epsilon_{\rm{F,ref}}$, $d=0.1R$, $g=2$, $r_{s}=1$, $R=0.5a_{B}r_{s}$, and $m=m_e$. Here, $r_{s}=1/\sqrt{\pi n_{e}a_{B}^2}$ is the Wigner-Seitz density parameter. Note, that in Fig.~\ref{fig:spectrum} the energy spectrum is only shown in the zeroth order in $d$, because in the final results of our calculations the single-particle eigenenergies enter only in the zeroth order.

At zero magnetic field [see Fig.~\ref{fig:spectrum}(a)] each subband - except the subbands with $n=0$, which are doubly
degenerate - has a fourfold degeneracy due to spin degeneracy and rotational invariance. For the chosen parameters, two degenerate pairs of subbands, namely a pair denoted by $n=0,\sigma=\downarrow,\Lambda=\pm1$ and a pair denoted by
$n=0,\sigma=\uparrow,\Lambda=\pm1$, are occupied as shown in Fig.~\ref{fig:spectrum}(a).

A finite magnetic field lifts the spin as well as the angular momentum degeneracy. With increasing strength of the magnetic field the energies of the two subbands with $\sigma=\uparrow$ start to increase towards the Fermi level, while the energies of the two subbands denoted by $\sigma=\downarrow$ decrease. This results in the occupation of four, five, and six non-degenerate bands, as shown in Figs.~\ref{fig:spectrum}(b), (c), and (d), respectively. Further increasing of the magnetic field strength leads to the inversion of the order of the ($n=0$, $\sigma=\uparrow$) and
($n=-1$, $\sigma=\downarrow$) subbands [compare Figs.~\ref{fig:spectrum}(d) and (e)] and, eventually, to the
depopulation of the ($n=0$, $\sigma=\uparrow$) subbands [see Figs.~\ref{fig:spectrum}(f)].

\subsection{Effects of Coulomb interaction}

We now include the effects of the electron-electron interaction. Electrons in the same tube interact via the intratube potentials
\begin{equation}
V_{\rm{out/in}}\left(l,q\right)=\frac{e^2}{\tilde{\epsilon}}I_{l}\left(|q|R\pm|q|\frac{d}{2}\right)K_{l}\left(|q|R\pm|q|\frac{d}{2}\right),
\end{equation}
depending on the tube both electrons are located in. On the other hand, the intertube Coulomb interaction between electrons from two different tubes is given by
\begin{equation}
V_{\rm d}\left(l,q\right)=\frac{e^2}{\tilde{\epsilon}}I_{l}\left(|q|R-|q|\frac{d}{2}\right)K_{l}\left(|q|R+|q|\frac{d}{2}\right).
\end{equation}
In these expressions, $I_{l}(x)$ and $K_{l}(x)$ are the modified Bessel functions, while $l$ and $q$ denote the change of angular and linear momentum (along the $z$-axis), respectively. It is convenient to define linear combinations
\begin{equation}
V_{\rm out}^{\pm}\left(l,q\right)=\frac{1}{2}\left[V_{\rm out}\left(l,q\right)\pm V_{\rm d}\left(l,q\right)\right],
\end{equation}
and
\begin{equation}
V_{\rm in}^{\pm}\left(l,q\right)=\frac{1}{2}\left[V_{\rm in}\left(l,q\right)\pm V_{\rm d}\left(l,q\right)\right].
\end{equation}
Then, the electron-electron interaction can be written as
\begin{equation}
\begin{array}{l}
\hat{H}'=\\
\frac{1}{2\pi L}\sum\limits_{l,q}\left\{\left[V_{\rm out}^{-}\left(l,q\right)+V_{\rm in}^{-}\left(l,q\right)\right]\hat{S}_{z}\left(l,q\right)\hat{S}_{z}\left(-l,-q\right)\right.\\
+\left.\frac{1}{4}\left[V_{\rm out}^{+}\left(l,q\right)+V_{\rm in}^{+}\left(l,q\right)\right]\hat{n}\left(l,q\right)\hat{n}\left(-l,-q\right)\right.\\
+\frac{1}{2}\left[V_{\rm out}^{-}\left(l,q\right)-V_{\rm in}^{-}\left(l,q\right)\right]\bigl[\hat{S}_{z}\left(l,q\right)\hat{n}\left(-l,-q\right)\\
\left.+\hat{n}\left(l,q\right)\hat{S}_{z}\left(-l,-q\right)\bigl]\right\}\\
-\frac{1}{2\pi L}\sum\limits_{l,q}\left[V_{\rm out}^{-}\left(l,q\right)-V_{\rm in}^{-}\left(l,q\right)\right]\hat{S}_{z},
\end{array}\label{InteractionHamiltonian2ndTTTT}
\end{equation}
where
\begin{equation}
\hat{n}\left(l,q\right)=\sum\limits_{n,k,\sigma,\Lambda}\hat{a}^{\dagger}_{n,k,\sigma,\Lambda}\hat{a}_{n+q,k+l,\sigma,\Lambda}\label{DensityOperator}
\end{equation}
is the local density operator. The complete Hamiltonian describing our system is
\begin{equation}
\hat{H}=\hat{H}_{0}+\hat{H}_{t}+\hat{H}',
\end{equation}
which comprises single-electron, tunneling, as well as Coulomb coupling terms.

\section{Pseudospin resonance}
\label{PR}

Our goal is to investigate the pseudospin resonance, which is an analog to the ferromagnetic
resonance,\cite{Gilbert2004:IEEETM,Hankiewicz2007:PRB} and how the resonance is affected by the electron-electron interaction. In what follows, zero temperature is considered. In the uniform case, that is, for zero momentum and angular momentum transfer, the pseudospin response function is, in the linear response regime,
\begin{equation}
\chi\left(\omega\right)=\frac{1}{2\pi L}\Braket{\Braket{\hat{S}_{z},\hat{S}_{z}}}_{\omega},
\end{equation}
where the brackets denote the Kubo product,
\begin{equation}
\braket{\braket{\hat{A},\hat{B}}}_{\omega}=-i\int\limits_{0}^{\infty}d\tau\quad e^{i\left(\omega+i\epsilon\right)\tau}\Bra{0}[\hat{A}(\tau),\hat{B}(0)]\Ket{0}.\label{KuboProduct}
\end{equation}
Since $\hat{S}_{z}$ measures the difference between the number of electrons in the outer and inner tubes, the pseudospin resonance describes collective oscillations of the particle densities between the tubes. This collective mode can be induced either by an external electric potential $V_{\rm{ext}}\left(t\right)$ applied between the tubes (as is also the case in flat-bilayer systems) or by fluctuations $B_{fl}\left(t\right)$ in the coaxial magnetic field (or by applying an oscillating, coaxial auxiliary magnetic field). The latter case is dealt with in the model by replacing the constant magnetic field amplitude $B$ with $B+B_{fl}(t)$ and treating the arising linear term containing $B_{fl}(t)$ as an external perturbation and neglecting the higher order terms. Using linear response theory\cite{Fetter:1980,Giuliani:2005} the fluctuations of the pseudospin expectation value of the system due to those perturbations can be calculated from the following expressions:
\begin{equation}
\delta\braket{\hat{S}_{z}\left(\omega\right)}=-\frac{2e}{\hbar}\chi\left(\omega\right)V_{ext}\left(\omega\right)\label{PseudospinResponseElectric}
\end{equation}
and
\begin{equation}
\delta\braket{\hat{S}_{z}\left(\omega\right)}=\frac{e}{m}\frac{\Phi}{\Phi_{0}}\frac{d}{R}\chi\left(\omega\right)B_{fl}\left(\omega\right),\label{PseudospinResponseMagnetic}
\end{equation}
where $V_{\rm{ext}}\left(\omega\right)$ and $B_{fl}(\omega)$ are the Fourier transforms of $V_{\rm{ext}}\left(t\right)$ and $B_{fl}(t)$. By comparing Eqs.~(\ref{PseudospinResponseElectric}) and~(\ref{PseudospinResponseMagnetic}) one can see that
\begin{equation}
V_B(\omega)\equiv-\frac{\hbar}{2m}\frac{d}{R}\frac{\Phi}{\Phi_0}B_{fl}(\omega)
\end{equation}
acts as an effective field which has the same effect as an external electric potential.

The pseudospin resonance is given by the condition
\begin{equation}
\operatorname{Re}\left[\chi^{-1}\left(\omega_{\rm res}\right)\right]=0,
\end{equation}
from which the resonance frequency $\omega_{\rm res}$ will be extracted.

As suggested by Eq.~(\ref{PseudospinResponseMagnetic}), in coaxial tubes an auxiliary magnetic flux piercing the system leads to fluctuations of the pseudospin expectation value and can therefore be used as an alternative control parameter for inducing collective oscillations.

\section{Perturbation theory}
\label{PT}

We will work in the limit $d\ll R$ and introduce the dimensionless intertube distance $\xi_{d}=d/R$ as an expansion
parameter. To calculate the resonance frequency in powers of $\xi_d$, a slightly modified version of the perturbation theory
developed in Ref.~\onlinecite{Abedinpour2007:PRL} is applied. The perturbation scheme from Ref.~\onlinecite{Abedinpour2007:PRL} has been modified to account for the cylindrical geometry considered
here. First, $\chi\left(\omega\right)$ is systematically decomposed into ground state expectation values and correlation
functions, and an exact equation for $\chi\left(\omega\right)$, analogous to Eq.~(3) in Ref.~\onlinecite{Abedinpour2007:PRL}, is derived. This scheme is rather elaborate and we refer to the Appendix \ref{append} [see Eq.~(\ref{ExactFormula})] for more details. The perturbation theory is based on the expansions
\begin{equation}
V_{\rm out}^{-}\left(l,q\right)=\frac{e^2|q|}{\tilde{\epsilon}}I_{l}'\left(|q|R\right)K_{l}\left(|q|R\right)d+\mathcal{O}(\xi_{d}^2)
\end{equation}
and
\begin{equation}
V_{\rm in}^{-}\left(l,q\right)=-\frac{e^2|q|}{\tilde{\epsilon}}I_{l}\left(|q|R\right)K_{l}'\left(|q|R\right)d+\mathcal{O}(\xi_{d}^2),
\end{equation}
which vanish in the zeroth order and can be considered small perturbations for $\xi_{d}=d/R\ll 1$. To calculate the correlation functions and expectation values, it is convenient to switch from the Heisenberg to the interaction picture and base the perturbation scheme on the perturbation $\hat{H}_{\rm{per}}$, which consists of those parts of $\hat{H}'$ that contain at least one factor $V_{\rm{out}}^{-}\left(l,q\right)$ or $V_{\rm{in}}^{-}\left(l,q\right)$. The unperturbed Hamiltonian $\hat{H}_{\rm{unper}}$ is then given by $\hat{H}$, with $V_{\rm{out}}^{-}\left(l,q\right)$ and
$V_{\rm{in}}^{-}\left(l,q\right)$ set to zero. In the following, we will restrict ourselves to the high-density limit, where we can neglect the density-density coupling in $\hat{H}'$ (which is not part of $\hat{H}_{\rm{per}}$) and use the Hamiltonian $\hat{H}_{\rm{unper}}=\hat{H}_{0}+\hat{H}_{t}$ and its ground state as our reference system.

The strategy is to expand $\chi\left(\omega\right)$ in powers of $\xi_d$ (see Appendix \ref{append} for details). The expansion series of $\chi\left(\omega\right)$ is then inverted and expanded again in powers of $\xi_d$ up to the same order as $\chi\left(\omega\right)$ was. After that, the zeros of this new power series are computed as a function of the intertube distance and expanded once again , which provides the expansion series of the resonance frequency in powers of $\xi_d$. As an important example, the pseudospin resonance frequency will be calculated in the following section.

\section{Pseudospin resonance frequency}
\label{PRF}

An approximate expression for the resonance frequency can be obtained by expanding the pseudospin response function up to the first order in $\xi_d$. In such a case the evaluation of $\chi\left(\omega\right)$ [see Eq.~(\ref{ExactFormula})] is
greatly simplified (see Appendix \ref{append} for details) and one obtains the following approximate relation,
\begin{equation}
\begin{aligned}
\chi\left(\omega\right)=&\frac{\Delta}{\hbar\Omega_{\omega}^2}\left(\mathcal{M}^{(0)}_{x}+\mathcal{M}^{(1)}_{x}\right)+\frac{2\Delta^2}{\hbar^3\Omega_{\omega}^4}\frac{1}{\left(2\pi L\right)^2}\\
&\times\sum\limits_{l,q}\left[V_{\rm out}^{-}\left(l,q\right)+V_{\rm in}^{-}\left(l,q\right)\right]^{(1)}f^{(0)}\left(l,q\right)\\
&+\mathcal{O}(\xi_{d}^2),
\end{aligned}\label{PseudospinResponse1stOrder}
\end{equation}
where
\begin{equation}\label{Omega-def}
\Omega_{\omega}=\sqrt{\omega^2-\frac{\Delta^2}{\hbar^2}}
\end{equation}
and
\begin{equation}
\mathcal{M}_{x}=\frac{1}{2\pi L}\left.\bra{0}\hat{S}_{x}\ket{0}\right|_{t=0}.
\end{equation}
Expanding $\mathcal{M}_{x}$ in powers of $\xi_d$ one obtains in the zeroth order
\begin{equation}
\mathcal{M}_{x}^{(0)}=\frac{1}{2\pi L}\sum\limits_{n,k,\sigma}\frac{n_{n,k,\sigma,-1}^{(0)}-n_{n,k,\sigma,+1}^{(0)}}{2},
\end{equation}
with
\begin{equation}\label{n-distribution}
n_{n,k,\sigma,\Lambda}^{(0)}=\Theta\left(\omega_{F}^{(0)}-\omega_{n,k,\sigma,\Lambda}^{(0)}\right),
\end{equation}
where $\omega_{F}^{(0)}=\epsilon_{F}^{(0)}/\hbar$ [$\epsilon_{F}^{(0)}$ is the Fermi energy of the system in the
zeroth order in $\xi_d$], and $\omega_{n,k,\sigma,\Lambda}^{(0)}=\epsilon_{n,k,\sigma,\Lambda}^{(0)}/\hbar$. The correction $\mathcal{M}_{x}^{(1)}$ adding the first order terms of $\mathcal{M}_{x}$ is irrelevant for the expansion of $\chi^{-1}\left(\omega\right)$ up to the first order in $\xi_d$. In the equations above, and in what follows, the superscripts denote the respective order in $\xi_d$. Finally, we have introduced
\begin{equation}\label{f0-def}
\begin{aligned}
f^{(0)}\left(l,q\right)=&\delta_{l,0}\delta_{q,0}\left(2\pi L\mathcal{M}^{(0)}_{x}\right)^2\\
&+\frac{1}{2}\sum\limits_{n,k,\sigma,\Lambda}n^{(0)}_{n,k,\sigma,\Lambda}\left(1-n^{(0)}_{n+l,k+q,\sigma,\Lambda}\right)\\
&-\frac{1}{4}\sum\limits_{n,k,\sigma}\sum\limits_{\Lambda,\Sigma}n^{(0)}_{n,k,\sigma,\Lambda}\left(1-n^{(0)}_{n+l,k+q,\sigma,\Sigma}\right),
\end{aligned}
\end{equation}
in Eq.~(\ref{PseudospinResponse1stOrder}).

Next, the approximate pseudospin response function given in Eq.~(\ref{PseudospinResponse1stOrder}) is inverted and then expanded in powers of $\xi_d$. From the zero of this series the resonance frequency $\omega_{\rm res}$ can be calculated. After several algebraic manipulations, the resonance frequency reads as
\begin{equation}
\omega_{\rm
res}^2=\frac{\Delta^2}{\hbar^2}+\frac{4\Delta}{\hbar^2}\frac{e^2}{\tilde{\epsilon}}\frac{d}{R}\frac{\mathcal{M}^{(0)}_{x}
\left(\uparrow\right)\mathcal{M}^{(0)}_{x}\left(\downarrow\right)}{\mathcal{M}^{(0)}_{x}}+\mathcal{O}(\xi_{d}^2).
\label{ResonanceFinal}
\end{equation}
For the derivation of this expression, the Wronskian formula
\begin{equation}
I_{l}(x)K'_{l}(x)-I'_{l}(x)K_{l}(x)=-\frac{1}{x},
\end{equation}
for modified Bessel functions\cite{Abramowitz:1972} has been used to rewrite the potential differences in the first order in $\xi_d$:
\begin{equation}
\left[V_{\rm out}^{-}\left(l,q\right)+V_{\rm in}^{-}\left(l,q\right)\right]^{(1)}=\frac{e^2}{\tilde{\epsilon}}\frac{d}{R}
\end{equation}
In Eq.~(\ref{ResonanceFinal}) we have introduced the spin-resolved pseudomagnetizations $\mathcal{M}^{(0)}_{x}\left(\uparrow\right)$ and $\mathcal{M}^{(0)}_{x}\left(\downarrow\right)$, which are defined as
\begin{equation}\label{mx-l-def}
\mathcal{M}_{x}^{(0)}(\sigma)=\frac{1}{2\pi L}\sum\limits_{n,k}\frac{n_{n,k,\sigma,-1}^{(0)}-n_{n,k,\sigma,+1}^{(0)}}{2}.
\end{equation}
Recall that the pseudospin quantum number $\Lambda=\pm1$ refers to the bonding and antibonding eigenstates of the Hamiltonian $\hat{H}_{0}+\hat{H}_{t}$ with eigenenergies $\epsilon_{n,k,\sigma,\Lambda}$, given by Eq.~(\ref{speettt}). In Eqs.~(\ref{n-distribution}) and (\ref{mx-l-def}) the energy spectrum enters only in its zeroth order in $\xi_d$.

It follows from Eqs.~(\ref{n-distribution}) and (\ref{mx-l-def}) that the spin-resolved pseudomagnetization measures the difference between the number of bonding ($\Lambda=-1$) and antibonding ($\Lambda=+1$) states for a fixed spin quantum number $\sigma$. Therefore, the total pseudomagnetization is given by the sum
\begin{equation}
\mathcal{M}_{x}^{(0)}=\sum\limits_{\sigma=\uparrow/\downarrow}\mathcal{M}_{x}^{(0)}(\sigma).
\end{equation}

From Eq.~(\ref{ResonanceFinal}) one can see that the Coulomb interaction produces a shift of the resonance frequency with
respect to $\Delta/\hbar$, the resonance frequency one would have obtained within the single-particle picture. This opens the possibility of investigating many-body effects by measuring the frequency shift of the pseudospin resonance.

At finite magnetic fields, the shift in the resonance frequency, given by
\begin{equation}
\Omega^2=\omega_{\rm res}^2-\frac{\Delta^2}{\hbar^2},
\end{equation}
depends on the different spin populations of the subbands denoted by the pseudospin quantum numbers $\Lambda=\pm1$. This is
due to Zeeman splitting which lifts the spin degeneracy of the different subbands. The division into spin-resolved
pseudomagnetizations $\mathcal{M}^{(0)}_{x}\left(\uparrow\right)$ and $\mathcal{M}^{(0)}_{x}\left(\downarrow\right)$ is important if the Zeeman term is not small compared to $\Delta$.

\begin{figure}
\includegraphics*[width=8cm]{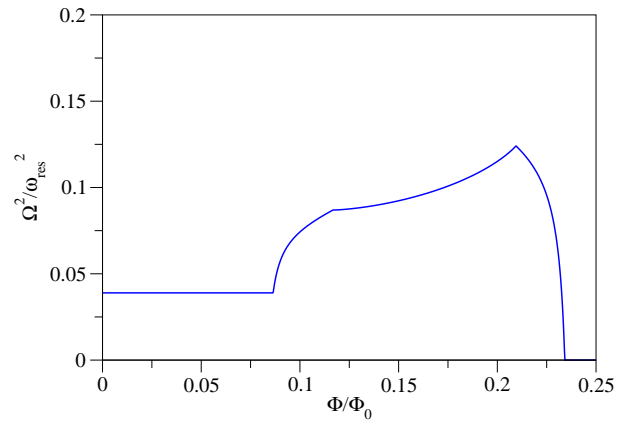}
\caption{Magnetic field dependence of the shift in the resonance frequency for a model system with parameters $\Delta=0.1\epsilon_{\rm F,ref}$, $d=0.1R$, $g=2$, $r_{s}=1$, $R=0.5a_{B}r_{s}$, and $m=m_e$.}\label{fig:Resonance}
\end{figure}

The magnetic field dependence of the shifted resonance frequency $\Omega$ is shown in Fig.~\ref{fig:Resonance} for a model system with parameters $\Delta=0.1\epsilon_{\rm F,ref}$, $d=0.1R$, $g=2$, $r_{s}=1$, $R=0.5a_{B}r_{s}$, and $m=m_e$, that is, the same parameters as in Fig.~\ref{fig:spectrum}. One can see that for small values of the magnetic field the shift of the resonance frequency is almost constant. This trend changes abruptly when the magnetic flux reaches the value 0.086 $\Phi_{0}$. At this point of non-analyticity the resonance frequency shift sharply starts to increase with the flux. For larger magnetic fields a pronounced peak of $\Omega^2$ develops. However, increasing the magnetic field even further results in a sharp drop of $\Omega^2$, and the shift of the resonance frequency vanishes for higher fields.

\begin{figure}
\includegraphics*[width=8cm]{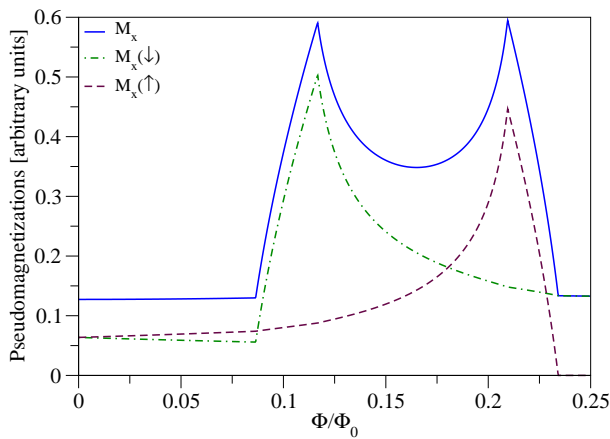}
\caption{(color on line). Magnetic field dependence of the spin-resolved and total pseudomagnetizations for the same generic
system as in Fig.~\ref{fig:Resonance}.}\label{fig:Pseudomagnetization}
\end{figure}

The behavior of the resonance frequency shift is determined, essentially, by the magnetic field dependence of the spin-resolved pseudomagnetizations [see Eq.~(\ref{ResonanceFinal})], which are shown in Fig.~\ref{fig:Pseudomagnetization}. As already mentioned, the spin-resolved pseudomagnetizations $\mathcal{M}^{(0)}_{x}\left(\uparrow\right)$ and $\mathcal{M}^{(0)}_{x}\left(\downarrow\right)$ measure the difference between the number of occupied bonding ($\Lambda=-1$) and antibonding ($\Lambda=+1$) states for up-spin and down-spin particles, respectively. Therefore, the magnetic field dependence of the spin-resolved pseudomagnetization can be qualitatively explained by analyzing the changes of the energy spectrum with the magnetic flux [see Fig.~\ref{fig:spectrum}].

In the limit of zero magnetic field both the bonding and antibonding states are spin degenerate [see Fig.~\ref{fig:spectrum}(a)]. Therefore, the single-particle energies no longer depend on the spin, implying
\begin{equation}
\mathcal{M}^{(0)}_{x}\left(\uparrow\right)=\mathcal{M}^{(0)}_{x}\left(\downarrow\right)=\frac{1}{2}\mathcal{M}^{(0)}_{x},
\end{equation}
which can be seen in Fig.~\ref{fig:Pseudomagnetization}. In such a case the resonance frequency shift given by Eq.~(\ref{ResonanceFinal}) reduces to
\begin{equation}
\Omega^2=\frac{\Delta}{\hbar^2}\frac{e^2}{\tilde{\epsilon}}\frac{d}{R}\mathcal{M}^{(0)}_{x}+\mathcal{O}(\xi_{d}^2).
\label{ResonanceFinalZeroField}
\end{equation}
Note that since the eigenstates are non-degenerate with respect to the quantum number $\Lambda$ [Fig.~\ref{fig:spectrum}(a)], the bonding and antibonding states are unequally occupied. This results in a finite value for the spin-resolved and total pseudomagnetizations at zero magnetic field.

The relation in Eq.~(\ref{ResonanceFinalZeroField}) is similar to that of the flat bilayer system examined in Ref.~\onlinecite{Abedinpour2007:PRL}. In fact, if we compare the different constituents to the frequency shift in Eq.~(\ref{ResonanceFinalZeroField}) and in Ref.~\onlinecite{Abedinpour2007:PRL} and their respective definitions, we can see that, in the first order in $\xi_d$, the structure of both expressions is the same.

At small magnetic fields the energy shift (and, therefore, the difference in occupation) of the bonding and antibonding states for both up- and down-spin particles changes smoothly with the field strength, leading only to small effects on the spin-resolved pseudomagnetizations. Nevertheless, one can still observe an increase (decrease) in $\mathcal{M}^{(0)}_{x}\left(\uparrow\right)$ [$\mathcal{M}^{(0)}_{x}\left(\downarrow\right)$] as the non-degenerate subbands with $n=0,\sigma=\uparrow,\Lambda=\pm1$ ($n=0,\sigma=\downarrow,\Lambda=\pm1$) shift up (down) towards (from) the Fermi level and the difference between their populations increases (decreases). The opposite behavior of $\mathcal{M}^{(0)}_{x}\left(\uparrow\right)$ and $\mathcal{M}^{(0)}_{x}\left(\downarrow\right)$ compensate each other, resulting in almost magnetic field-independent total pseudomagnetization and resonance frequency shift, in the region
$0\leq \Phi/\Phi_{0}<0.086$ [see Figs.~\ref{fig:Resonance} and \ref{fig:Pseudomagnetization}]. The situation changes drastically at $\Phi\approx0.086\Phi_{0}$, at which point a fifth subband, namely the subband with $n=-1,\sigma=\downarrow,\Lambda=-1$, becomes occupied [see Fig.~\ref{fig:spectrum}(c)]. This results in a sharp increase of $\mathcal{M}^{(0)}_{x}\left(\downarrow\right)$ because now there is an additional subband, which is rapidly populated and contributes to the $\Lambda=-1$ states in $\mathcal{M}^{(0)}_{x}\left(\downarrow\right)$, while there is only one band with $\Lambda=+1$ that contributes to $\mathcal{M}^{(0)}_{x}\left(\downarrow\right)$. At these magnetic fields, the spacing in the energy spectrum between states which differ only in their pseudospin quantum number $\Lambda$ is already much smaller than the spacing between states which differ only in their real spin quantum number $\sigma$. At $\Phi\approx0.117\Phi_{0}$, the subband with $n=-1,\sigma=\downarrow,\Lambda=+1$ becomes occupied [see Fig.~\ref{fig:spectrum}(d)], which consequently leads to a decay in the absolute value of $\mathcal{M}^{(0)}_{x}\left(\downarrow\right)$. This behavior continues until the subbands with $n=-1,\sigma=\downarrow,\Lambda=-1$ and $n=-1,\sigma=\downarrow,\Lambda=+1$, which are energetically very close to each other, lie far below the Fermi level [see Fig.~\ref{fig:spectrum}(f)], and the decay becomes less pronounced. If we look at the behavior of $\mathcal{M}^{(0)}_{x}\left(\uparrow\right)$ in Fig.~\ref{fig:Pseudomagnetization}, we can observe a steady increase, which becomes more steep as the minimum of the antibonding subband with $n=0,\sigma=\uparrow,\Lambda=+1$ moves closer to the Fermi level [see Figs.~\ref{fig:spectrum}(c) and (d)]. At $\Phi\approx0.21\Phi_{0}$ this subband is no longer occupied, while the bonding subband with $n=0,\sigma=\uparrow,\Lambda=-1$ is still populated [see Fig.~\ref{fig:spectrum}(e)]. This results in the peak of $\mathcal{M}^{(0)}_{x}\left(\uparrow\right)$, which can be seen clearly in Fig.~\ref{fig:Pseudomagnetization}. With even higher magnetic fields the remaining bonding subband with $n=0,\sigma=\uparrow,\Lambda=-1$ moves up and becomes less and less populated, leading to the decay of $\mathcal{M}^{(0)}_{x}\left(\uparrow\right)$ until the band minimum crosses the Fermi level (at $\Phi\approx0.234\Phi_{0}$). For $\Phi>0.234\Phi_{0}$ the spin-up subbands lie already above the Fermi level and only spin-down subbands are occupied [see Fig.~\ref{fig:spectrum}(f)]. Consequently, $\mathcal{M}^{(0)}_{x}\left(\uparrow\right)$ vanishes in this region [see Fig.~\ref{fig:Pseudomagnetization}].

\begin{figure}
\includegraphics*[width=8cm]{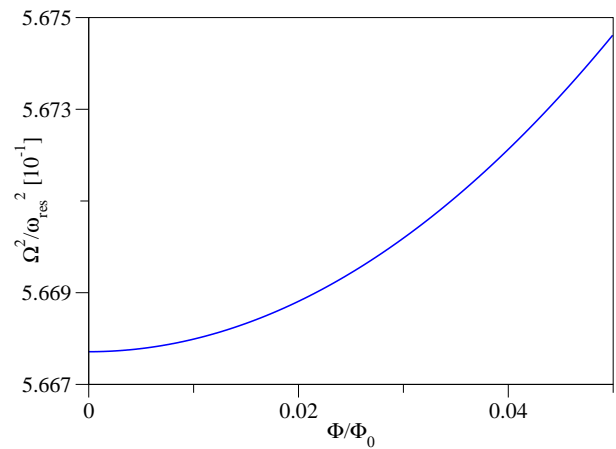}
\caption{Magnetic field dependence of the shift in the resonance frequency for a narrow model carbon nanotube system ($\epsilon_{r}=2.4$, $m=0.25m_{e}$, $g=2$, $R_{\rm{in}}=1.1$ nm, $R_{\rm{out}}=1.44$ nm, $n_{1}=9.1$ $\rm{nm}^{-1}$, and $\Delta=0.1\epsilon_{\rm F,ref}$)\label{fig:CNS}}
\end{figure}

Finally, the magnetic field dependence of the resonance frequency shift [see Fig.~\ref{fig:Resonance}] is determined by the interplay of the above discussed behaviors of both the spin-resolved and total pseudomagnetizations.

\begin{figure}
\includegraphics*[width=8cm]{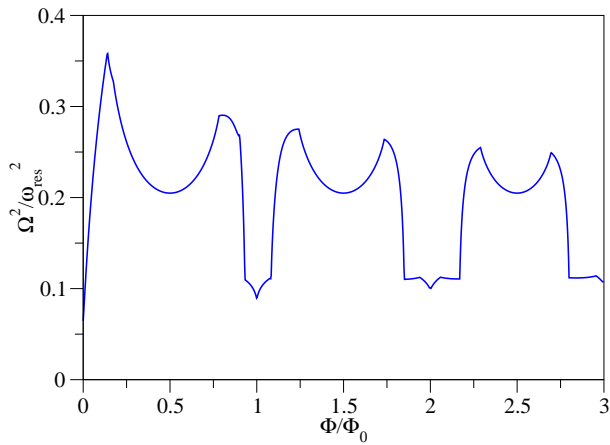}
\caption{Magnetic field dependence of the shift in the resonance frequency for an InGaAs/GaAs nanotube system ($a_B=9.8$ nm,
$g=-0.44$, $R=10$ nm, $d=1$ nm, $n=1\times10^{11}$ $\rm{cm}^{-2}$, and $\Delta=0.1\epsilon_{\rm F,ref}$)\label{fig:SCS}}
\end{figure}

In Figs.~\ref{fig:CNS} and \ref{fig:SCS} we show the magnetic field dependence of the resonance frequency shift for realistic model systems and for magnetic fluxes which correspond to magnetic fields of up to $40$ T. In Fig.~\ref{fig:CNS} the system parameters $\epsilon_{r}=2.4$, $m=0.25m_{e}$ and $g=2$ have been chosen to simulate a narrow carbon nanotube system with the radii $R_{\rm{in}}=1.1$ nm and $R_{\rm{out}}=1.44$ nm and the electron density per unit length $n_{1}=9.1$ $\rm{nm}^{-1}$.\cite{Lin1993:PRB,Lin1992:PRB,Dresselhaus:1987,Chauvet1995:PRB,Zhang2006:PRL}

In Fig.~\ref{fig:SCS} the parameters correspond to a model InGaAs/GaAs nanotube system ($a_B=9.8$ nm and $g=-0.44$) with an
average radius of $R=10$ nm, a distance $d=1$ nm between the two tubes, and an electron density $n=10^{11}$ $\rm{cm}^{-2}$.\cite{Weisbuch1977:PRB,Prinz2000:PE,Mendach2004:PE}

As in Figs.~\ref{fig:Resonance} and~\ref{fig:Pseudomagnetization}, the magnetic field dependence of the pseudomagnetizations and resonance frequency shift displayed in Figs.~\ref{fig:CNS} and \ref{fig:SCS} can be explained by analyzing how the single-particle energy spectrum evolves with the external magnetic field. The only difference is that in Figs.~\ref{fig:CNS} and~\ref{fig:SCS} more subbands are occupied than in Fig.~\ref{fig:Resonance}, which results in a more complicated structure of $\Omega^2$. Furthermore, the effect of Zeeman splitting is not as strong as in Fig.~\ref{fig:Resonance}, where the parameters were chosen for simplicity to illustrate Eq.~\ref{ResonanceFinal}. Therefore, one would have to go to unrealistically high magnetic fields to observe the vanishing of the resonance frequency shift. For magnetic fields below $40$ T one has not yet entered this regime for the systems shown in Figs.~\ref{fig:CNS} and \ref{fig:SCS}. There are several peaks for magnetic fields below $40$ T in the InGaAs/GaAs nanotube system (see Fig.~\ref{fig:SCS}). In the narrow carbon nanotube system, on the other hand, the resonance frequency shift does not vary much for the fields considered (see Fig.~\ref{fig:CNS}). Only at unrealistically high magnetic fields pronounced peaks would appear in this system and one would need even higher magnetic fields to observe the vanishing of the shift of the resonance frequency.

\section{Linewidth of the pseudospin resonance}
\label{LW}

Additional corrections to Eq.~(\ref{ResonanceFinal}) are obtained by expanding the pseudospin response function up to the second order in $\xi_d$, at which point a finite imaginary part of $\chi^{-1}\left(\omega\right)$ emerges. A non-vanishing
imaginary part implies damping and a finite lifetime of the mode. This means that the pole of $\chi\left(\omega\right)$ is -at least up to the second order in $\xi_d$- replaced by a roughly Lorentzian-shaped peak at the resonance frequency. The linewidth $\Gamma$ of the resonance can be obtained from the imaginary part of the zero of $\chi^{-1}\left(\omega\right)$ (in contrast to the zero of $\operatorname{Re}\left[\chi^{-1}\left(\omega\right)\right]$, which only yields the resonance frequency). This situation is the pseudospin analog to the finite linewidth of the ferromagnetic resonance induced by the Gilbert damping.

Using the shorthand notations $\textbf{u}=(n,k)$, $\textbf{v}=(l,q)$, and $\textbf{v}'=(l',q')$, the linewidth up to
second order in $\xi_d$ is given by
\begin{equation}\label{PSRDR}
\Gamma=\frac{\Delta}{\mathcal{M}_x^{(0)}\hbar^3}\left(\frac{e^2}{\tilde{\epsilon}}\frac{d}{R}\right)^2\lim\limits_{\omega\rightarrow\Delta/\hbar}\left[\frac{\mathcal{A}\left(\omega\right)}{\omega}+\frac{\mathcal{B}\left(\omega\right)}{\omega}\right]+\mathcal{O}(\xi_{d}^3)
\end{equation}
Here, $\mathcal{A}\left(\omega\right)$ and $\mathcal{B}\left(\omega\right)$ denote the sums
\begin{widetext}
\begin{equation}
\mathcal{A}\left(\omega\right)=\frac{\pi}{4}\frac{1}{\left(2\pi L\right)^3}\sum\limits_{\textbf{u},\sigma}\sum\limits_{\textbf{v},\textbf{v}'}\sum\limits_{\Lambda,\Sigma}\delta\left(\omega-\Omega_{\Lambda}(\textbf{v},\textbf{v}')\right)\left(1-n_{\textbf{u}+\textbf{v},\sigma,\Lambda}^{(0)}\right)\left(1-n_{\textbf{u}-\textbf{v}',-\sigma,\Sigma}^{(0)}\right)n_{\textbf{u},\sigma,\Sigma}^{(0)}n_{\textbf{u}+\textbf{v}-\textbf{v}',-\sigma,-\Lambda}^{(0)}
\end{equation}
and
\begin{equation}
\begin{aligned}
\mathcal{B}\left(\omega\right)=&\frac{\pi}{2}\frac{R^2}{\left(2\pi L\right)^3}\sum\limits_{\textbf{u},\textbf{v},\textbf{v}'}\sum\limits_{\sigma,\sigma',\Lambda}\mathcal{V}^2\left(\textbf{v}\right)\delta\left(\omega-\tilde{\Omega}(\textbf{v},\textbf{v}')\right)\left(1-n_{\textbf{u}+\textbf{v},\sigma,\Lambda}^{(0)}\right)\left(1-n_{\textbf{u}-\textbf{v}',\sigma',\Lambda}^{(0)}\right)n_{\textbf{u},\sigma,\Lambda}^{(0)}n_{\textbf{u}+\textbf{v}-\textbf{v}',\sigma',\Lambda}^{(0)}\\
&-\frac{\pi}{2}\frac{R^2}{\left(2\pi L\right)^3}\sum\limits_{\textbf{u},\textbf{v},\textbf{v}'}\sum\limits_{\sigma,\Lambda}\mathcal{V}\left(\textbf{v}\right)\mathcal{V}\left(\textbf{v}'\right)\delta\left(\omega-\tilde{\Omega}(\textbf{v},\textbf{v}')\right)\left(1-n_{\textbf{u}+\textbf{v},\sigma,\Lambda}^{(0)}\right)\left(1-n_{\textbf{u}-\textbf{v}',\sigma,\Lambda}^{(0)}\right)n_{\textbf{u},\sigma,\Lambda}^{(0)}n_{\textbf{u}+\textbf{v}-\textbf{v}',\sigma,\Lambda}^{(0)}\quad,
\end{aligned}
\end{equation}
\end{widetext}
where
\begin{equation}
\Omega_{\Lambda}(\textbf{v},\textbf{v}')=\frac{\hbar}{2mR^2}ll'+\frac{\hbar}{m}qq'+\frac{\Delta}{\hbar}\Lambda,
\end{equation}
\begin{equation}
\tilde{\Omega}(\textbf{v},\textbf{v}')=\frac{\hbar}{2mR^2}ll'+\frac{\hbar}{m}qq',
\end{equation}
and
\begin{equation}
\mathcal{V}\left(\textbf{v}\right)=|q|\left[I'_{l}\left(|q|R\right)K_{l}\left(|q|R\right)+I_{l}\left(|q|R\right)K'_{l}\left(|q|R\right)\right].
\end{equation}
The zeroth order band occupation numbers $n_{\textbf{u},\sigma,\Lambda}^{(0)}$ have already been introduced in Eq.~(\ref{n-distribution}). The physical origin of the damping can be understood by looking at the expressions for
$\mathcal{A}\left(\omega\right)$ and $\mathcal{B}\left(\omega\right)$. The mode loses energy by exciting two particle-hole pairs out of the single-particle spectrum. The process described by $\mathcal{A}\left(\omega\right)$ consists of an excitation where the pseudospin is conserved and another one where the pseudospin is flipped. Furthermore, the total momentum, the total angular momentum and the total spin are conserved during this double-excitation. The second process is given by $\mathcal{B}\left(\omega\right)$, which also describes a double-particle-hole excitation. Like before, the process is momentum-, angular momentum-, and spin-conserving. But in this case the process involves either only bonding or only antibonding states, and there is no pseudospin-flip. The process given by $\mathcal{A}\left(\omega\right)$ can also be found in the flat bilayer system,~\cite{Abedinpour2007:PRL} while $\mathcal{B}\left(\omega\right)$ is a manifestation of the
cylindrical system and the existence of two different intratube-potentials.

\begin{figure}
\includegraphics*[width=8cm]{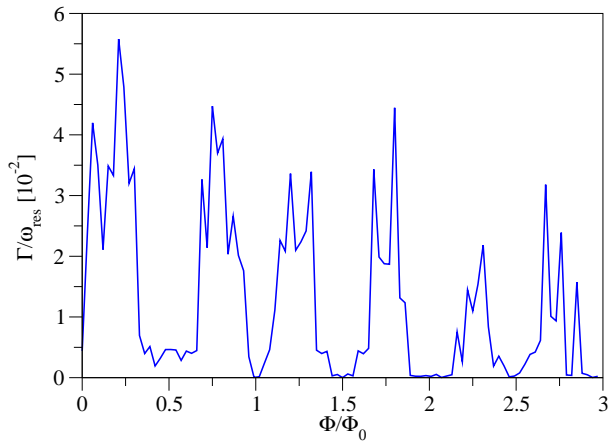}
\caption{Magnetic field dependence of the linewidth of the pseudospin resonance for an InGaAs/GaAs nanotube system ($a_B=9.8$ nm, $g=-0.44$, $R=10$ nm, $d=1$ nm, $L=1$ $\mu$m, $n=1\times10^{11}$ $\rm{cm}^{-2}$, and $\Delta=0.1\epsilon_{F}^{0}$)\label{fig:SCSDR}}
\end{figure}

The magnetic field dependence of the resonance decay rate computed from Eq.~(\ref{PSRDR}) is shown in Fig.~\ref{fig:SCSDR} for the case of a semiconductor system with the same parameters as in Fig.~\ref{fig:SCS}. Additionally, we chose a finite length $L=1$ $\mu {\rm m}$ for the nanotubes to speed up the time-consuming numerical evaluation of Eq.~(\ref{PSRDR}) by discretizing the linear momenta in Eq.~(\ref{PSRDR}). The decay rate exhibits a strong dependence on the magnetic field, with several pronounced peaks at which the resonance linewidth is enhanced and regions where the damping is almost suppressed (that is, where $\Gamma$ is close to zero) and the pseudospin resonance becomes very sharp. This interesting behavior makes the external magnetic field, which is an experimentally tunable parameter, attractive for the controlled switching of the damping of the pseudospin oscillations.

\section{Conclusion}
\label{Cs}

We have considered a cylindrical bilayer system consisting of two coaxial tubes. To account for the presence of two tubes, we have introduced a new quantum number which describes the two-level system. This two-level system was then interpreted in terms of pseudospin-$1/2$ dynamics. We have incorporated tunneling between the two tubes and calculated the energy spectrum of the single-particle Hamiltonian. The inclusion of tunneling between the two tubes made it possible to find a pseudospin analog to the ferromagnetic resonance. Taking into account the electron-electron interaction, the pseudospin resonance frequency has been calculated up to the first order in the intertube distance by using a perturbative scheme, which has been developed for a flat bilayer system and appears to be superior to the RPA.~\cite{Abedinpour2007:PRL} Due to the Coulomb interaction there is a shift in the resonance frequency, which is also dependent on the coaxial magnetic field. This dependence results in pronounced peaks of the resonance frequency shift at certain magnetic fields. The shift, however, disappears at higher magnetic fields. The damping effects induced by the Coulomb electron-electron interaction on the pseudospin resonance have been investigated by computing the linewidth of the resonance. The linewidth exhibits a strong dependence on the magnetic field, with a multi-peak structure. Apart from the peaks, where the decay rate is enhanced, in some ranges of the magnetic filed strength the damping of the pseudospin oscillations is almost suppressed.

\acknowledgments{We thank F. Baruffa, S. Konschuh, and J. Najjar for valuable discussions. This work was supported by the Deutsche Forschungsgemeinschaft via GRK 638 and SFB 689.}

\appendix

\section{}
\label{append}

Here, we present the expression for $\chi\left(\omega\right)$ on which the perturbation scheme is based. The application of the Kubo product formulas
\begin{equation}
\Braket{\Braket{\hat{A},\hat{B}}}_{\omega}=\frac{1}{\omega}\bra{0}[\hat{A}(0),\hat{B}(0)]\ket{0}+\frac{i}{\omega}\Braket{\Braket{\partial_{t}A,\hat{B}}}_{\omega},\label{KuboProductIdentity1}
\end{equation}
and
\begin{equation}
\Braket{\Braket{\hat{A},\hat{B}}}_{\omega}=\frac{1}{\omega}\bra{0}[\hat{A}(0),\hat{B}(0)]\ket{0}-\frac{i}{\omega}\Braket{\Braket{\hat{A},\partial_{t}\hat{B}}}_{\omega},\label{KuboProductIdentity2}
\end{equation}
which can be verified by partial integration, leads to the following expression for the pseudospin response function
\begin{equation}
\begin{aligned}
&\chi\left(\omega\right)=\\
&\frac{\Delta}{\hbar\Omega_{\omega}^2}\mathcal{M}_{x}-\frac{i\Delta}{2\pi L\hbar^2\Omega_{\omega}^2}\mathcal{C}_{1}
\left(\omega\right)\\
&+\frac{\Delta}{\hbar^2\Omega_{\omega}^4}\frac{1}{\left(2\pi
L\right)^2}\sum\limits_{\textbf{v}}\mathcal{V}_{1}(\textbf{v})
\left(-i\omega g\left(\textbf{v}\right)+\frac{2\Delta}{\hbar}f\left(\textbf{v}\right)\right)\\
&-\frac{i\Delta\omega}{\hbar^2\Omega_{\omega}^4}\frac{1}{\left(2\pi
L\right)^2}
\sum\limits_{\textbf{v}}\mathcal{V}_{2}(\textbf{v})\left(h_{y}\left(\textbf{v}\right)-2\pi L\mathcal{M}_{y}\right)\\
&-\frac{\Delta^2}{\hbar^3\Omega_{\omega}^4}\frac{1}{\left(2\pi
L\right)^2}\sum\limits_{\textbf{v}}\mathcal{V}_{2}(\textbf{v})
\left(h_{z}\left(\textbf{v}\right)-2\pi L\mathcal{M}_{z}\right)\\
&+\frac{\Delta^2}{\hbar^4\Omega_{\omega}^4}\frac{1}{\left(2\pi
L\right)^3}
\sum\limits_{\textbf{v}}\sum\limits_{\textbf{v}'}\mathcal{V}_{1}(\textbf{v})\mathcal{V}_{1}(\textbf{v}')
\mathcal{L}_{0}\left(\textbf{v},\textbf{v}',\omega\right)\\
&+\frac{\Delta^2}{\hbar^4\Omega_{\omega}^4}\frac{1}{\left(2\pi
L\right)^3}\sum\limits_{\textbf{v}}
\sum\limits_{\textbf{v}'}\mathcal{V}_{1}(\textbf{v})\mathcal{V}_{2}(\textbf{v}')\mathcal{L}_{1}
\left(\textbf{v},\textbf{v}',\omega\right)\\
&+\frac{\Delta^2}{\hbar^4\Omega_{\omega}^4}\frac{1}{\left(2\pi
L\right)^3}\sum\limits_{\textbf{v}}
\sum\limits_{\textbf{v}'}\mathcal{V}_{2}(\textbf{v})\mathcal{V}_{1}(\textbf{v}')\tilde{\mathcal{L}}_{1}
\left(\textbf{v},\textbf{v}',\omega\right)\\
&+\frac{\Delta^2}{\hbar^4\Omega_{\omega}^4}\frac{1}{\left(2\pi
L\right)^3}\sum\limits_{\textbf{v}}
\sum\limits_{\textbf{v}'}\mathcal{V}_{2}(\textbf{v})\mathcal{V}_{2}(\textbf{v}')\tilde{\mathcal{L}}_{0}
\left(\textbf{v},\textbf{v}',\omega\right)\\
&-\frac{i\Delta^2}{\hbar^4\Omega_{\omega}^4}\frac{1}{\left(2\pi
L\right)^2}\sum\limits_{\textbf{v}}
\left(\mathcal{V}_{1}(\textbf{v})\mathcal{C}_{2}\left(\textbf{v},\omega\right)+\mathcal{V}_{2}(\textbf{v})
\mathcal{C}_{3}\left(\textbf{v},\omega\right)\right),
\end{aligned}\label{ExactFormula}
\end{equation}
where we have introduced the quantities
\begin{equation}
\mathcal{V}_{1}(\textbf{v})=\left[V_{\rm out}^{-}\left(\textbf{v}\right)+V_{\rm in}^{-}\left(\textbf{v}\right)\right],
\end{equation}
\begin{equation}
\mathcal{V}_{2}(\textbf{v})=\left[V_{\rm out}^{-}\left(\textbf{v}\right)-V_{\rm in}^{-}\left(\textbf{v}\right)\right],
\end{equation}
\begin{equation}
\mathcal{M}_{i}=\frac{1}{2\pi L}\left.\bra{0}\hat{S}_{i}\ket{0}\right|_{t=0},\label{DefPseudospinQuantitiesFirst}
\end{equation}
\begin{equation}
f\left(\textbf{v}\right)=\left.\bra{0}\hat{S}_{x}\left(\textbf{v}\right)\hat{S}_{x}\left(-\textbf{v}\right)-
\hat{S}_{z}\left(\textbf{v}\right)\hat{S}_{z}\left(-\textbf{v}\right)\ket{0}\right|_{t=0},
\end{equation}
\begin{equation}
g\left(\textbf{v}\right)=\left.\bra{0}\hat{S}_{y}\left(\textbf{v}\right)\hat{S}_{z}\left(-\textbf{v}\right)+
\hat{S}_{z}\left(\textbf{v}\right)\hat{S}_{y}\left(-\textbf{v}\right)\ket{0}\right|_{t=0},
\end{equation}
\begin{equation}
h_{i}\left(\textbf{v}\right)=\frac{1}{2}\left.\bra{0}\hat{S}_{i}\left(\textbf{v}\right)\hat{n}
\left(-\textbf{v}\right)+\hat{n}\left(\textbf{v}\right)\hat{S}_{i}\left(-\textbf{v}\right)\ket{0}\right|_{t=0},
\end{equation}
\begin{equation}
\mathcal{L}_{0}\left(\textbf{v},\textbf{v}',\omega\right)=\left\langle
\left\langle\hat{\mathcal{S}}_{xz}(\textbf{v}),\hat{\mathcal{S}}_{xz}(\textbf{v}')\right\rangle\right\rangle_{\omega},
\end{equation}
\begin{equation}
\mathcal{L}_{1}\left(\textbf{v},\textbf{v}',\omega\right)=\left\langle\left\langle
\hat{\mathcal{S}}_{xz}(\textbf{v}),\hat{\mathcal{Q}}(\textbf{v}')\right\rangle\right\rangle_{\omega},
\end{equation}
\begin{equation}
\tilde{\mathcal{L}}_{0}\left(\textbf{v},\textbf{v}',\omega\right)=\left\langle
\left\langle\hat{\mathcal{Q}}(\textbf{v}),\hat{\mathcal{Q}}(\textbf{v}')\right\rangle\right\rangle_{\omega},
\end{equation}
\begin{equation}
\tilde{\mathcal{L}}_{1}\left(\textbf{v},\textbf{v}',\omega\right)=\left\langle
\left\langle\hat{\mathcal{Q}}(\textbf{v}),\hat{\mathcal{S}}_{xz}(\textbf{v}')\right\rangle\right\rangle_{\omega},
\end{equation}
\begin{equation}
\mathcal{C}_{1}\left(\omega\right)=\Braket{\Braket{\left[\hat{S}_{y},\hat{H}_{0}\right],
\hat{S}_{z}}}_{\omega},
\end{equation}
\begin{equation}
\mathcal{C}_{2}\left(\textbf{v},\omega\right)=\left\langle\left\langle\hat{\mathcal{S}}_{xz}(\textbf{v}),
\left[\hat{S}_{y},\hat{H}_{0}\right]\right\rangle\right\rangle_{\omega},
\end{equation}
\begin{equation}
\mathcal{C}_{3}\left(\textbf{v},\omega\right)=\left\langle\left\langle\hat{\mathcal{Q}}(\textbf{v}),
\left[\hat{S}_{y},\hat{H}_{0}\right]\right\rangle\right\rangle_{\omega},
\end{equation}
\begin{equation}
\hat{\mathcal{S}}_{xz}(\textbf{v})=\hat{S}_{x}\left(\textbf{v}\right)\hat{S}_{z}\left(-\textbf{v}\right)+
\hat{S}_{z}\left(\textbf{v}\right)\hat{S}_{x}\left(-\textbf{v}\right),
\end{equation}
\begin{equation}
\hat{\mathcal{Q}}(\textbf{v})=\frac{1}{2}\left[\hat{S}_{x}\left(\textbf{v}\right)\hat{n}\left(-\textbf{v}\right)+
\hat{n}\left(\textbf{v}\right)\hat{S}_{x}\left(-\textbf{v}\right)-2\hat{S}_{x}\right],
\label{DefPseudospinQuantitiesLast}
\end{equation}
and the shorthand notations $\textbf{v}=(l,q)$ and $\textbf{v}'=(l',q')$. This equation is analogous to Eq.~(3) in Ref.~\onlinecite{Abedinpour2007:PRL}, but due to the intrinsic dependence of the single-particle eigenstates of $\hat{H}_{0}$ on the interlayer difference and different interlayer Coulomb potentials in each tube, there are additional terms. If this intrinsic interlayer dependence were absent, both intralayer Coulomb interactions would be identical, which means every term that contains the factor $\left[V_{\rm{out}}^{-}\left(l,q\right)-V_{\rm{in}}^{-}\left(l,q\right)\right]$ would vanish. Furthermore, $\hat{H}_{0}$ would no longer depend on $d$ and would commute with $\hat{S}_{y}$. In this case, Eq.~(\ref{ExactFormula}) would reduce to Eq.~(3) in Ref.~\onlinecite{Abedinpour2007:PRL}.

To obtain the resonance frequency in the first order, we need to expand the response function in Eq.~(\ref{ExactFormula}) up to first order in $\xi_d$. In such a case, we do not have to consider the terms containing the products $\mathcal{V}_{i}(\textbf{v})\mathcal{V}_{j}(\textbf{v}')$ $(i,j=1,2)$ because they are already of second order or higher in $\xi_d$. The situation is further simplified by looking at the commutator of $\hat{H}_{0}$ and $\hat{S}_{y}$, which also vanishes in the zeroth order in $\xi_d$ and thus can only contribute to first or higher order of the pseudospin response function. This implies, however, that products of this commutator with $\mathcal{V}_{i}(\textbf{v})$ $(i=1,2)$ do not add to the pseudospin response function up to the first order. As a result, many terms in Eq.~(\ref{ExactFormula}) can be omitted in the first order. The calculation of the remaining terms shows that only $f(l,q)$ and the pseudomagnetization $\mathcal{M}_{x}$ do not vanish in the zeroth order in $\xi_d$. Moreover, $\Braket{\Braket{\left[\hat{S}_{y},\hat{H}_{0}\right],\hat{S}_{z}}}_{\omega}$ also vanishes in zeroth and first orders in $\xi_d$, which means that this term cannot contribute to the pseudospin response function up to the first order. Consequently, we are left with the formula given in Eq.~(\ref{ResonanceFinal}), which is a simplified expression of the response function accurate up to the first order in $\xi_d$.

For the linewidth of the pseudospin resonance, one has to calculate $\chi\left(\omega\right)$ in at least second order in $\xi_d$, which means no term in Eq.~(\ref{ExactFormula}) can be omitted.

\bibliographystyle{apsrev}


\end{document}